\documentclass[a4paper, 11pt]{article}
\usepackage{ascmac}
\usepackage{mathrsfs}
\usepackage{latexsym}
\usepackage{amsmath}
\usepackage{amssymb}
\usepackage{amsthm}

\usepackage[hdivide={30mm,,30mm}, vdivide={20mm,,25mm}, nohead]{geometry}

 \makeatletter
    
    \@addtoreset{equation}{section}
  \makeatother

\newtheorem{Thm}{Theorem}[section]
\newtheorem{Lem}{Lemma}[section]
\newtheorem{Prop}{Proposition}[section]

\newtheorem{Rem}{Remark}[section]

\DeclareMathOperator*{\slim}{s-lim}
\begin{document}
\title{Absence of wave operators for one-dimensional
\\
quantum walks
}
\author{Kazuyuki Wada\\ Department of General Science and Education,
\\
 {National Institute of Technology, Hachinohe college.}\\\ Hachinohe, 039-1192, Japan.\vspace{3mm}\\ E-mail address: kazuyuki.dawa@gmail.com  
}
\date{}
\maketitle
\begin{abstract}
We show that there exist pairs of two time evolution operators which do not have wave operators in a context of one-dimensional discrete time quantum walks. As a consequence, the borderline between short range type and long range type is decided. 
\end{abstract}

\section{Introduction} 
We consider a discrete time quantum walk on $\mathbb{Z}$. Let $\mathcal{H}:=l^{2}(\mathbb{Z};\mathbb{C}^{2})$ be a Hilbert space and $U:=SC$ be the unitary time evolution operator of quantum walks. Here $S$ is a shift operator and $C$ is a coin operator. The axiom of quantum walks is introduced in [12] and the classification of one-dimensional quantum walks is considered in [7].
Quantum walks have been introduced as a quantum counter part of classical random walks [1,4]. It is known that the behavior of quantum walks is different from classical random walks. One of differences appears in a weak limit theorem which is regarded as a quantum walk version of central limit theorem. Konno firstly proved this theorem if a coin operator is position independent of $\mathbb{Z}$ [6]. An interesting consequence is that the shape of a limit distribution in quantum walks is different from the normal distribution which can be derived from central limit theorem for classical random walks. After that, several researchers extend his result [see e.g. 3,10,11,13]. According to [3], the asymptotic velocity operator plays important roles to get weak limit theorems. Moreover in [3], the explicit form of the asymptotic velocity operator of position independent quantum walks is established through discrete Fourier transforms. 

In this paper, we mainly consider a position dependent quantum walk. Namely, $C$ is a multiplication operator by a unitary matrix $C(x)\in U(2)$, $x\in\mathbb{Z}$. If $C$ depends on a position $x\in\mathbb{Z}$, it is difficult to know the form of asymptotic velocity operator since the discrete Fourier transform does not work. To overcome this difficulty, Suzuki introduced the discrete time wave operator for quantum walks in [13]. Suppose that there exist $ C_{0}\in U(2)$ and constants $\epsilon, \kappa>0$ such that
\begin{equation}
\|C(x)-C_{0}\|_{\mathcal{B}(\mathbb{C}^{2})}\le\kappa(1+|x|)^{-1-\epsilon},\hspace{5mm}x\in\mathbb{Z},
\end{equation}
where $\|\cdot\|_{\mathcal{B}(\mathbb{C}^{2})}$ is the operator norm on $\mathbb{C}^{2}$. 
We set $U_{0}:=SC_{0}$. In [10, 11,13], the above type condition is called the short range type condition. Under this condition, following wave operators exist and are complete:
\begin{equation}\label{wave}
W_{\pm}(U, U_{0}):=\slim_{t\rightarrow\pm\infty}U^{-t}U_{0}^{t}\Pi_{\text{ac}}(U_{0}),
\end{equation}
where $\slim$ denotes the strong limit and $\Pi_{\text{ac}}(U_{0})$ denotes the orthogonal projection onto the absolutely continuous subspace of $U_{0}$. Moreover in [13], Suzuki introduced the asymptotic velocity operator by using above wave operators and derived the weak limit theorem for position dependent cases. This result is extended in several models (see e.g. [10, 11]).

 The main problem in this paper is the existence or non-existence of wave operators if $C$ and $C_{0}$ satisfy
\begin{equation}
\|C(x)-C_{0}\|_{\mathcal{B}(\mathbb{C}^{2})}\le \kappa(1+ |x|)^{-\gamma},\hspace{5mm}x\in\mathbb{Z},
\end{equation}
for some $\kappa>0$ and $\gamma\in(0, 1]$. Then we say that $C$ and $C_{0}$ satisfy the long range type condition. In a context of Schr\"{o}dinger operators, it is known that if a potential slowly converges to 0 at infinity, then the wave operator does not exist in general [2, 5, 8, 9]. From this fact, it is expected that similar situations occur in a context of quantum walks. Consequently, this expectation is true. In other words, there exist examples of $U$ and $U_{0}$ such that wave operators do not exist. Therefore we can conclude that the borderline between short range type and long range type is $\gamma=1$.
Some results related to non-existence of wave operators are known in a context of Schr\"{o}dinger operators [2, 5, 8, 9]. In these cases, we can expect the borderline between short range type condition and long range type condition from the large time behavior of a classical orbit of a particle. For these heuristic arguments, see e.g. [5]. 

To show the non-existence of wave operators, we employ the argument introduced by Ozawa [8]. We need careful treatments since the time evolution is discrete. If $C_{0}$ is diagonal, then the proof is quite simple since the motion of a quantum walker is simple (Remark 3.1). On the other hand, if $C_{0}$ is not diagonal, then the proof is complicated. Let $Q$ be the position operator on $\mathbb{Z}$, $Q(t)$ be the Heisenberg operator of $Q$ by $U_{0}$ and $V_{0}$ be the asymptotic velocity operator of $U_{0}$. Roughly speaking, the weak limit theorem says that   $Q(t)/t$ converges to $V_{0}$ as $t\rightarrow\infty$ in a suitable sense. To complete the proof, this fact is crucial. Key lemmas related to the weak limit theorem are stated in Lemma 3.1 and Lemma 3.3.

Contents of this paper are as follows. In section 2, we review notation for quantum walks and state the main result. In section 3 we give a proof of the main result.

\section{Main result}
In this section we review notation for quantum walks and state the main result in this paper. The Hilbert space is given by
\begin{equation}
\mathcal{H}:=l^{2}(\mathbb{Z};\mathbb{C}^{2})=\Big\{\Psi:\mathbb{Z}\rightarrow\mathbb{C}^{2}\Big| \displaystyle\sum_{x\in\mathbb{Z}}\|\Psi(x)\|^{2}_{\mathbb{C}^{2}}<\infty\Big\},
\end{equation}
where $\|\cdot\|_{\mathbb{C}^{2}}$ is the norm on $\mathbb{C}^{2}$. We denote its inner product and norm by $\langle\cdot, \cdot \rangle_{\mathcal{H}}$ (linear in the right vector)  and $\|\cdot\|_{\mathcal{H}}$, respectively. If there is no danger of confusion, then we omit the subscript $\mathcal{H}$ of them. We introduce the following dense subspace of $\mathcal{H}$:
\begin{equation}
\mathcal{H}_{0}:=\{\phi\in\mathcal{H}| \exists N\in\mathbb{N}\text{ such that }\phi(x)=0\text{ for all }|x|\ge N\}.
\end{equation}
Next we introduce two unitary operators $U$ and $U_{0}$. For $\Psi\in\mathcal{H}$, the shift operator $S$ is defined by
\begin{equation}
(S\Psi)(x):=\begin{bmatrix}\Psi^{(1)}(x+1) \\ \Psi^{(2)}(x-1)\end{bmatrix}, \hspace{5mm}x\in\mathbb{Z}.
\end{equation}
For $C_{0}\in U(2)$ and $\gamma>0$, we introduce the following coin operator $C$:
\begin{equation}\label{coin}
(C\Psi)(x):=C(x)\Psi(x), \hspace{5mm}C(x):=e^{i(1+|x|)^{-\gamma}}C_{0},\hspace{5mm}x\in\mathbb{Z},
\end{equation}
where $i$ is the imaginary unit.
Throughout in this paper, we identify $C_{0}$ as a unitary operator on $\mathcal{H}$ such that  $(C_{0}\Psi)(x)=C_{0}\Psi(x)$, $x\in\mathbb{Z}$. We set $U:=SC$ and $U_{0}:=SC_{0}$.

Let $\|\cdot\|_{\mathcal{B}(\mathbb{C}^{2})}$ be the operator norm on $\mathbb{C}^{2}$. For any $x\in\mathbb{Z}$, it is seen that
\begin{equation}
\displaystyle\frac{1}{2}(1+|x|)^{-\gamma}\le\|C(x)-C_{0}\|_{\mathcal{B}(\mathbb{C}^{2})}\le (1+|x|)^{-\gamma}.
\end{equation}

For any $C_{0}\in U(2)$, $C_{0}$ has a form of
\begin{equation}
C_{0}=\begin{bmatrix}a & b\\ -e^{i\delta} b^{\ast} & e^{i\delta} a^{\ast}\end{bmatrix},
\end{equation}
where $e^{i\delta}$ $(\delta\in[0, 2\pi))$ is the determinant of $C_{0}$ and for $z\in\mathbb{C}$, $z^{\ast}$ denotes the complex conjugate of $z$. We note that $a$ and $b$ satisfy $|a|^{2}+|b|^{2}=1$. 
\begin{Rem}\normalfont
In this paper, our goal is to find the example of $(U, U_{0})$ such that wave operators do not exist. Thus we only consider the coin operator introduced by $(\ref{coin})$.
\end{Rem}
Let $A$ be a unitary or self-adjoint operator on $\mathcal{H}$. The sets $\sigma(A)$, $\sigma_{\text{p}}(A)$, $\sigma_{\text{c}}(A)$ and $\sigma_{\text{ac}}(A)$ are called spectrum, pure point spectrum, continuous spectrum and absolutely continuous spectrum of $A$, respectively.
For spectral properties of $U_{0}$, following facts are known:
\begin{Prop}\normalfont[10, Lemma 4.1 and Proposition 4.5]
\begin{enumerate}
\item If $|a|=1$, then $U_{0}$ has purely absolutely continuous spectrum and $\sigma(U_{0})=\sigma_{\text{ac}}(U_{0})=\{e^{i\tau}|\tau\in[0, 2\pi)\}$.
\item If $0<|a|<1$, $U_{0}$ has purely absolutely continuous spectrum and
\begin{equation*}
\sigma(U_{0})=\sigma_{\text{ac}}(U_{0})=\{e^{i\tau}|\tau\in[\delta/2+\tau, \pi+\delta/2-\tau]\cup[\pi+\delta/2+\tau, 2\pi+\delta/2-\tau]\},
\end{equation*}
where $\theta:=\arccos |a|$.
\item If $a=0$, then $U_{0}$ has pure point spectrum and  $\sigma(U_{0})=\sigma_{\text{p}}(U_{0})=\{ie^{i\delta/2}, -ie^{i\delta/2}\}$.
\end{enumerate}
\end{Prop}
We are interested in cases $|a|=1$ and $0<|a|<1$. The main result is as follows:

\begin{Thm}\normalfont
For any $a\in\mathbb{C}$ with $0<|a|\le 1$ and $\gamma\in(0, 1]$, $\displaystyle\slim_{t\rightarrow\pm\infty}U^{-t}U_{0}^{t}$
does not exist.
\end{Thm}
From Theorem 2.1, we can conclude that the borderline between short range type and the long range type is $\gamma=1$. 
\section{Proof of Theorem 2.1}
In this section we prove Theorem 2.1. First of all, we assume that $|a|=1$. Then $b=0$ since $C_{0}$ is unitary matrix. Thus $C_{0}$ has a form of
\begin{equation*}
C_{0}=\begin{bmatrix}a & 0 \\ 0 & a^{\ast}\end{bmatrix}.
\end{equation*}
\begin{Rem}\normalfont
Since $C_{0}$ is diagonal, the motion of a quantum walker by $U_{0}$ is as follows:
\begin{enumerate}
\item An element of a set $\Big\{\Psi\in\mathcal{H}\Big|\Psi(x)=\begin{bmatrix}\Psi^{(1)}(x) \\ 0\end{bmatrix},\hspace{2mm}x\in\mathbb{Z}\Big\}$ only moves to left.
\\
\item An element of a set $\Big\{\Psi\in\mathcal{H}\Big|\Psi(x)=\begin{bmatrix}0 \\ \Psi^{(2)}(x) \end{bmatrix},\hspace{2mm}x\in\mathbb{Z}\Big\}$ only moves to right.
\end{enumerate}
\end{Rem}
\begin{proof}[Proof of Theorem 2 ($|a|=1$)]
We only consider the case $t\rightarrow\infty$. The other case is also proven by the similar manner. We take $\phi\in\mathcal{H}_{0}$. Then there exists $M\in\mathbb{N}$ such that $\phi(x)=0$ if $|x|>M$.
Suppose that $\phi_{+}:=\lim_{t\rightarrow\infty}U^{-t}U_{0}^{t}\phi$ exists. 
Since $\|U^{t}\phi_{+}-U_{0}^{t}\phi\|=\|\phi_{+}-U^{-t}U_{0}^{t}\phi\|\rightarrow 0$ (as $t\rightarrow\infty$), we can take $N\in\mathbb{N}$ so that $\|U^{t}\phi_{+}-U_{0}^{t}\phi\|\le \|\phi\|^{2}/4$ if $t\ge N$. We set $W(t):=U^{-t}U_{0}^{t}$. Then it follows that
\begin{equation}
W(t_{2})-W(t_{1})=\displaystyle\sum_{t=t_{1}+1}^{t_{2}}U^{-t}(U_{0}-U)U_{0}^{t-1},\hspace{5mm}t_{2}>t_{1}>0.
\end{equation}
For $t_{2}>t_{1}>\text{max}\{2M, N\}$+1, we have
\begin{equation*}
\begin{aligned}
&\hspace{3mm}\text{Im}\langle \{W(t_{2})-W(t_{1})\}\phi, \phi_{+}\rangle
\\
&=\displaystyle\sum_{t=t_{1}+1}^{t_{2}}\text{Im}\langle (U_{0}-U)U_{0}^{t-1}\phi, U_{0}^{t}\phi\rangle+\displaystyle\sum_{t=t_{1}+1}^{t_{2}}\text{Im}\langle (U_{0}-U)U_{0}^{t-1}\phi, U^{t}\phi_{+}-U_{0}^{t}\phi\rangle
\\
&=\displaystyle\sum_{t=t_{1}+1}^{t_{2}}\text{Im}\langle (C_{0}-C)U_{0}^{t-1}\phi, C_{0}U_{0}^{t-1}\phi\rangle+\displaystyle\sum_{t=t_{1}+1}^{t_{2}}\text{Im}\langle (U_{0}-U)U_{0}^{t-1}\phi, U^{t}\phi_{+}-U_{0}^{t}\phi\rangle
\\
&=\displaystyle\sum_{t=t_{1}+1}^{t_{2}}\displaystyle\sum_{x\in\mathbb{Z}}\sin(1+|x|)^{-\gamma}\|(U_{0}^{t-1}\phi)(x)\|_{\mathbb{C}^{2}}^{2}+\displaystyle\sum_{t=t_{1}+1}^{t_{2}}\text{Im}\langle (U_{0}-U)U_{0}^{t-1}\phi, U^{t}\phi_{+}-U_{0}^{t}\phi\rangle,
\end{aligned}
\end{equation*}
where $\text{Im}z$ is the imaginary part of $z\in\mathbb{C}$. By $t\ge 2M+1$ and Remark 3.1, the intersection of a support of $U_{0}^{t-1}\phi$ and $\{-t+M, \cdots, t-M\}$ is empty. Thus we have
\begin{equation*}
\begin{aligned} 
&\hspace{3mm}\text{Im}\langle \{W(t_{2})-W(t_{1})\}\phi, \phi_{+}\rangle
\\
&\ge\displaystyle\frac{\|\phi\|^{2}}{2}\displaystyle\sum_{t=t_{1}+1}^{t_{2}}(1+t-M)^{-\gamma}-\displaystyle\sum_{t=t_{1}+1}^{t_{2}}\|(C_{0}-C)U_{0}^{t-1}\phi\|\|U^{t}\phi_{+}-U_{0}^{t}\phi\|
\\
&\ge\displaystyle\frac{\|\phi\|^{2}}{2}\displaystyle\sum_{t=t_{1}+1}^{t_{2}}(1+t-M)^{-\gamma}-\displaystyle\frac{\|\phi\|^{2}}{4}\displaystyle\sum_{t=t_{1}+1}^{t_{2}}(1+t-M)^{-\gamma}
\\
&=\displaystyle\frac{\|\phi\|^{2}}{4}\displaystyle\sum_{t=t_{1}+1}^{t_{2}}(1+t-M)^{-\gamma} \rightarrow\infty \hspace{2mm}(\text{as}\hspace{2mm}t_{2}\rightarrow\infty).
\end{aligned}
\end{equation*}
On the other hand, $\text{Im}\langle (W(t_{2})-W(t_{1}))\phi, \phi_{+}\rangle$ is bounded by $2\|\phi\|^{2}$. This is a contradiction. 
\end{proof}

Hereafter, we assume that $0<|a|<1$. In this case, we need more preparations. 
We set the Hilbert space $\mathcal{K}:=L^{2}([0, 2\pi), \frac{dk}{2\pi};\mathbb{C}^{2})$ and $\mathcal{F}:\mathcal{H}\rightarrow\mathcal{K}$ be the discrete Fourier transform which is the unitary operator defined as the unique continuous extension of the following operator:
\begin{equation}
(\mathcal{F}\phi)(k):=\displaystyle\sum_{x\in\mathbb{Z}}\phi(x)e^{-ikx},\hspace{5mm}\phi\in\mathcal{H}_{0},\hspace{5mm}k\in[0, 2\pi).
\end{equation}
In what follows, we denotes the image of the discrete Fourier transform of $\phi\in\mathcal{H}$ by $\hat{\phi}$.
We define $\hat{U}_{0}:=\mathcal{F}U_{0}\mathcal{F}^{-1}$. $\hat{U}_{0}$ is decomposable and it follows that
\begin{equation}
(\hat{U}_{0}f)(k)=\hat{U}_{0}(k)f(k)\hspace{3mm}\text{with}\hspace{3mm}\hat{U}_{0}(k)=\begin{bmatrix}e^{ik} & 0 \\ 0 & e^{-ik}\end{bmatrix}C_{0},\hspace{3mm}f\in\mathcal{K},\hspace{3mm}\text{ a.e. }k\in[0, 2\pi).
\end{equation} 
We denote an eigenvalue and a correspond normalized eigenvector by $\lambda_{j}(k)$ and $u_{j}(k)$ $(j=1, 2)$, respectively. We set
\begin{equation*}
\begin{aligned}
\tau(k)&:=|a|\cos(k+\arg(a)-\delta/2),
\\
\eta(k)&:=\sqrt{1-\tau(k)^{2}},
\\
\zeta(k)&:=|a|\sin(k+\arg(a)-\delta/2),
\end{aligned}
\end{equation*}where $\arg(z)\in[0, 2\pi)$ is the argument of $z\in\mathbb{C}$. It is known that $\lambda_{j}(k)$ and $u_{j}(k)$ can be expressed as
\begin{equation}\label{ev}
\lambda_{j}(k)=e^{i\delta}\{\tau(k)+i(-1)^{j-1}\eta(k)\},\hspace{3mm}u_{j}(k)=\displaystyle\frac{\sqrt{\eta(k)+(-1)^{j-1}\zeta(k)}}{|b|\sqrt{2\eta(k)}}\begin{bmatrix}i|b|e^{i(k+\arg(b)-\delta/2)} \\ \zeta(k)+(-1)^{j}\eta(k)\end{bmatrix},\hspace{3mm}(j=1,2).
\end{equation}
For details, see e.g.[10]. From (\ref{ev}), $\hat{U}_{0}(k)$ is expressed as
\begin{equation}\label{u}
\hat{U}_{0}(k)=\displaystyle\sum_{j=1, 2}\lambda_{j}(k)\langle u_{j}(k),\cdot\rangle_{\mathbb{C}^{2}}u_{j}(k), \hspace{5mm}k\in [0, 2\pi),
\end{equation}
\begin{Rem}\normalfont
It is seen that $\lambda_{j}(k)$ is a $2\pi$ periodic $C^{\infty}$ function in the variable $k$ and $\mathbb{C}^{2}$-valued function $u_{j}(k)$ is also a $2\pi$ periodic $C^{\infty}$ function in the variable $k$. Moreover following quantities are finite:
\begin{equation*}
\sup_{0\le k<2\pi}|\lambda_{j}'(k)|, |\lambda_{j}''(k)|<\infty,\hspace{5mm}\sup_{0\le k<2\pi}\|u_{j}(k)\|_{\mathbb{C}^{2}}<\infty,\hspace{5mm}(j=1, 2), 
\end{equation*}
where $\lambda_{j}'(k)$ and $u_{j}'(k)$ are derivatives of $\lambda_{j}(k)$ and $u_{j}(k)$, respectively and $\lambda_{j}''(k)$ is the second derivative of $\lambda_{j}(k)$. These facts are used in latter lemmas.
\end{Rem}

Next we introduce the asymptotic velocity operator of $U_{0}$. We denote it by $V_{0}$ and is given by  
\begin{equation*}
(\widehat{V_{0}f})(k):=\displaystyle\sum_{j=1,2}\displaystyle\frac{i\lambda'_{j}(k)}{\lambda_{j}(k)}\langle u_{j}(k),\hat{f}(k)\rangle_{\mathbb{C}^{2}}u_{j}(k),\hspace{3mm}f\in\mathcal{H},\hspace{3mm}k\in[0, 2\pi).
\end{equation*}
We note that $V_{0}$ is bounded and self-adjoint on $\mathcal{H}$. 
\begin{Prop}\normalfont[10, Lemma 4.2 (b)]
If $0<|a|<1$, then $\sigma_{\text{p}}(V_{0})=\phi$ and $\sigma(V_{0})=\sigma_{\text{c}}(V_{0})=[-|a|, |a|]$.
\end{Prop}
Let us denote a subspace of vectors $\phi\in \mathcal{D}$ whose discrete Fourier transform $\hat{\phi}$ is differentiable in a variable $k$ and
\begin{equation*}
\displaystyle\sup_{k\in[0, 2\pi)}\Big\|\displaystyle\frac{\text{d}}{\text{d}k}\hat{\phi}(k)\Big\|<\infty.
\end{equation*}
Let $Q$ be a position operator defined by
\begin{equation*}
\text{dom}(Q):=\Big\{\phi\in\mathcal{H}\Big|\displaystyle\sum_{x\in\mathbb{Z}}x^{2}\|\phi(x)\|_{\mathbb{C}^{2}}^{2}<\infty\Big\},\hspace{2mm}(Q\phi)(x):=x\phi(x),\hspace{2mm}x\in\mathbb{Z}, \hspace{2mm}\phi\in D(Q),
\end{equation*}
where $\text{dom}(Q)$ is the domain of $Q$. We set $Q(t):=U_{0}^{-t}QU_{0}^{t}$ and $D:=\mathcal{F}Q\mathcal{F}^{-1}$. For $\phi\in \mathcal{D}$, it is seen that $(D\hat{\phi})(k)=i\displaystyle\frac{\text{d}}{\text{d}k}\hat{\phi}(k)$.
Following lemmas are important in our analysis:
\begin{Lem}\normalfont
For any $\phi\in \mathcal{D}$, there exists a constant $\kappa_{1}>0$ which is independent of $t$ such that
\begin{equation*}
\Big\|\displaystyle\frac{Q(t)}{t}\phi-V_{0}\phi\Big\|\le \kappa_{1} t^{-1},\hspace{5mm}t\in\mathbb{Z}\setminus \{0\}.
\end{equation*}
\end{Lem}
\begin{proof} Although it is established in the proof of Theorem 4.1 in [13], we give a proof for completeness. By the discrete Fourier transforms, it is seen that
\begin{equation*}
\Big\|\Big(\displaystyle\frac{Q(t)}{t}-V_{0}\Big)\phi\Big\|^{2}
=\displaystyle\int_{0}^{2\pi}\Big\|\hat{U}(k)^{-t}\displaystyle\frac{i}{t}\displaystyle\frac{\text{d}}{\text{d}k}\Big(\hat{U}(k)^{t}\hat{\phi}(k)\Big)-\displaystyle\sum_{j=1, 2}\displaystyle\frac{i\lambda_{j}'(k)}{\lambda_{j}(k)}\langle u_{j}(k), \hat{\phi}(k)\rangle_{\mathbb{C}^{2}}u_{j}(k)\Big\|_{\mathbb{C}^{2}}\displaystyle\frac{\text{d}k}{2\pi}.
\end{equation*}
From (\ref{u}), it is seen that
\begin{equation*}
\begin{aligned}
\hat{U}(k)^{-t}\displaystyle\frac{i}{t}\displaystyle\frac{\text{d}}{\text{d}k}\big\{\hat{U}(k)^{t}\hat{\phi}(k)\big\}&=\hat{U}(k)^{-t}\displaystyle\frac{i}{t}\displaystyle\frac{\text{d}}{\text{d}k}\Big(\displaystyle\sum_{j=1, 2}\lambda_{j}(k)^{t}\langle u_{j}(k), \hat{\phi}(k)\rangle_{\mathbb{C}^{2}}u_{j}(k)\Big)
\\
&=\displaystyle\sum_{j=1, 2}\displaystyle\frac{i\lambda_{j}'(k)}{\lambda_{j}(k)}\langle u_{j}(k), \hat{\phi}(k)\rangle_{\mathbb{C}^{2}}u_{j}(k)\\
&\hspace{10mm}+\displaystyle\frac{i}{t}\hat{U}(k)^{-t}\displaystyle\sum_{j=1, 2}\lambda_{j}(k)^{t}\displaystyle\frac{\text{d}}{\text{d}k}\Big(\langle u_{j}(k), \hat{\phi}(k)\rangle_{\mathbb{C}^{2}}u_{j}(k)\Big).
\end{aligned}
\end{equation*}
Therefore we have
\begin{equation*}
\Big\|\Big(\displaystyle\frac{Q(t)}{t}-V_{0}\Big)\phi\Big\|^{2}
=\displaystyle\frac{1}{t^{2}}\displaystyle\int_{0}^{2\pi}\Big\|\displaystyle\sum_{j=1, 2}\lambda_{j}(k)^{t}\displaystyle\frac{\text{d}}{\text{d}k}\Big(\langle u_{j}(k), \hat{\phi}(k)\rangle_{\mathbb{C}^{2}}u_{j}(k)\Big)\Big\|^{2}_{\mathbb{C}^{2}}\displaystyle\frac{\text{d}k}{2\pi}.
\end{equation*}
By the definition of $\mathcal{D}$ and Remark 3.2, we have 
\begin{equation*}
\displaystyle\sup_{0\le k<2\pi}\displaystyle\Big\|\frac{\text{d}}{\text{d}k}\Big(\langle u_{j}(k), \hat{\phi}(k)\rangle_{\mathbb{C}^{2}}u_{j}(k)\Big)\Big\|_{\mathbb{C}^{2}}<\infty.
\end{equation*}
Thus we have the desired inequality.
\end{proof}
\begin{Lem}\normalfont
For any $\phi\in\mathcal{D}$, there exist positive constants $L_{1}$ and $L_{2}$ such that for any $z\in\mathbb{C}$ with $\text{Im}z\neq0$, 
\begin{equation*}
\Big\|\Big(V_{0}-\displaystyle\frac{Q(t)}{t}\Big)(z-V_{0})^{-1}\phi\Big\|\le \Big(L_{1}|\text{Im}z|^{-1}+L_{2}|\text{Im}z|^{-2}|\Big)|t|^{-1},\hspace{5mm}t\in\mathbb{Z}\setminus\{0\}.
\end{equation*}
\end{Lem}
\begin{proof}
By the discrete Fourier transform, it is seen that
\begin{equation}
\begin{aligned}
&\Big\|\Big(V_{0}-\displaystyle\frac{Q(t)}{t}\Big)(z-V_{0})^{-1}\phi\Big\|^{2}
\\
&=\displaystyle\frac{1}{t^{2}}\displaystyle\int_{0}^{2\pi}\Big\|\displaystyle\sum_{j=1, 2}\lambda_{j}(k)^{t}\displaystyle\frac{\text{d}}{\text{d}k}\Big(\Big(z-\displaystyle\frac{i\lambda_{j}'(k)}{\lambda_{j}(k)}\Big)^{-1}\langle u_{j}(k), \hat{\phi}(k)\rangle_{\mathbb{C}^{2}}u_{j}(k)\Big)\Big\|_{\mathbb{C}^{2}}^{2}\displaystyle\frac{\text{d}k}{2\pi}.
\end{aligned}
\end{equation}
A direct calculation yields that
\begin{equation*}
\begin{aligned}
&\displaystyle\frac{\text{d}}{\text{d}k}\Big(\Big(z-\displaystyle\frac{i\lambda_{j}'(k)}{\lambda_{j}(k)}\Big)^{-1}\langle u_{j}(k), \hat{\phi}(k)\rangle_{\mathbb{C}^{2}}u_{j}(k)\Big)
\\
&=-\Big(z-\displaystyle\frac{i\lambda'_{j}(k)}{\lambda_{j}(k)}\Big)^{-2}\displaystyle\frac{\text{d}}{\text{d}k}\Big(\displaystyle\frac{i\lambda'_{j}(k)}{\lambda_{j}(k)}\Big)\langle u_{j}(k), \hat{\phi}(k)\rangle_{\mathbb{C}^{2}}u_{j}(k)
\\
&\hspace{15mm}+\Big(z-\displaystyle\frac{i\lambda'_{j}(k)}{\lambda_{j}(k)}\Big)^{-1}\displaystyle\frac{\text{d}}{\text{d}k}\Big(\langle u_{j}(k), \hat{\phi}(k)\rangle u_{j}(k)\Big)
\\
&=-i\Big(z-\displaystyle\frac{i\lambda'_{j}(k)}{\lambda_{j}(k)}\Big)^{-2}\displaystyle\frac{\lambda_{j}''(k)\lambda_{j}(k)-\big(\lambda_{j}'(k)\big)^{2}}{\lambda_{j}(k)^{2}}\langle u_{j}(k), \hat{\phi}(k)\rangle_{\mathbb{C}^{2}}u_{j}(k)
\\
&\hspace{15mm}+\Big(z-\displaystyle\frac{i\lambda'_{j}(k)}{\lambda_{j}(k)}\Big)^{-1}\displaystyle\frac{\text{d}}{\text{d}k}\Big(\langle u_{j}(k), \hat{\phi}(k)\rangle u_{j}(k)\Big).
\end{aligned}
\end{equation*}
By the definition of $\mathcal{D}$ and Remark 3.2, there exist constants $C_{1}$ and $C_{2}$ such that
\begin{equation}
\displaystyle\sup_{0\le k<2\pi}\Big\|\displaystyle\frac{\text{d}}{\text{d}k}\Big(\Big(z-\displaystyle\frac{i\lambda_{j}'(k)}{\lambda_{j}(k)}\Big)^{-1}\langle u_{j}(k), \hat{\phi}(k)\rangle_{\mathbb{C}^{2}}u_{j}(k)\Big)\Big\|_{\mathbb{C}^{2}}^{2}\le C_{1}|\text{Im}z|^{-1}+C_{2}|\text{Im}z|^{-2}.
\end{equation}
From (3.6) and (3.7), we have the desired result.
\end{proof}
We introduce the following set of functions:
\begin{equation*}
C_{0}^{\infty}(\mathbb{R}):=\{f\in C^{\infty}(\mathbb{R})| f \text{ has a compact support}\}.
\end{equation*}
\begin{Lem}\normalfont For any $G\in C_{0}^{\infty}(\mathbb{R})$ and $\phi\in\mathcal{D}$, there exists a constant $\kappa_{2}>0$ which is independent of $t$ such that
\begin{equation*}
\Big\|G\Big(\displaystyle\frac{Q(t)}{t}\Big)\phi-G(V_{0})\phi\Big\|\le \kappa_{2}t^{-1},\hspace{5mm}t\in\mathbb{Z}\setminus\{0\}.
\end{equation*}
\end{Lem}
\begin{proof}
We apply the Helffer-Sj\"{o}srand formula[2]. For a self-adjoint operator $A$, it follows that
\begin{equation}\label{HS}
G(A)=\displaystyle\frac{1}{2\pi i}\displaystyle\int_{\mathbb{C}}(\overline{\partial}\tilde{G})(z)(z-A)^{-1}\text{d}z\text{d}\overline{z},
\end{equation}
where $z=x+iy$, $\overline{\partial}=\frac{1}{2}(\partial_{x}+i\partial_{y})$ and $\tilde{G}$ is the almost analytic extension of $G$ which satisfies following properties:
\begin{enumerate}
\item $\tilde{G}(x)=G(x)$ if $x\in\mathbb{R}$,
\item $\tilde{G}$ is infinitely many differentiable in $x$ and $y$,
\item A support of $\tilde{G}$ is compact in $\mathbb{C}$,
\item For any $N\in\mathbb{N}$, there exists a constant $C_{N}$ such that $|\overline{\partial}\tilde{G}(z)|\le C_{N}|\text{Im}z|^{N}$.
\end{enumerate}
We note that the integral on the right hand side of (\ref{HS}) is taken in the sense of operator norm topology. By using it, we have
\begin{equation*}
\begin{aligned}
&\hspace{5mm}\Big\|G\Big(\displaystyle\frac{Q(t)}{t}\Big)\phi-G(V_{0})\phi\Big\|
\\
&\le\displaystyle\frac{1}{2\pi }\displaystyle\int_{\mathbb{C}}|(\overline{\partial}\tilde{G})(z)|\Big\|\Big(z-\displaystyle\frac{Q(t)}{t}\Big)^{-1}\Big\|\Big\|\Big(V_{0}-\displaystyle\frac{Q(t)}{t}\Big)(z-V_{0})^{-1}\phi\Big\|\text{d}z\text{d}\overline{z}
\\
&\le \displaystyle\frac{1}{2\pi }\displaystyle\int_{\mathbb{C}}|(\overline{\partial}\tilde{G})(z)||\text{Im}z|^{-1}\Big\|\Big(V_{0}-\displaystyle\frac{Q(t)}{t}\Big)(z-V_{0})^{-1}\phi\Big\|\text{d}z\text{d}\overline{z}.
\end{aligned}
\end{equation*}
From Lemma 3.2, there exist positive constants $L_{1}$ and $L_{2}$ such that
\begin{equation*}
\Big\|\Big(V_{0}-\displaystyle\frac{Q(t)}{t}\Big)(z-V_{0})\phi\Big\|\le \Big(L_{1}|\text{Im}z|^{-1}+L_{2}|\text{Im}z|^{-2}\Big)|t|^{-1},\hspace{5mm}t\neq0.
\end{equation*} 
By the property of $\tilde{G}$, there exists a constant $C_{3}>0$ such that $|(\overline{\partial}\tilde{G})(z)|\le C_{3}|\text{Im}z|^{3}$. Since the support of $\tilde{G}$ is compact, we have 
\begin{equation*}
\begin{aligned}
\Big\|G\Big(\displaystyle\frac{Q(t)}{t}\Big)\phi-G(V_{0})\phi\Big\|
&\le \displaystyle\frac{1}{2\pi}\displaystyle\int_{\mathbb{C}}|\overline{\partial}\tilde{G}(z)||\text{Im}z|^{-1}\big(L_{1}|\text{Im}z|^{-1}+L_{2}|\text{Im}z|^{-2}\big)|t|^{-1}\text{d}z\text{d}\overline{z}
\\
&\le |t|^{-1}\times\underbrace{\displaystyle\frac{C_{3}}{2\pi}\displaystyle\int_{\text{supp}\tilde{G}}\big(L_{1}|\text{Im}z|+L_{2}\big)\text{d}z\text{d}\overline{z}}_{:=\kappa_{2}},
\end{aligned}
\end{equation*}
where $\text{supp}\tilde{G}$ is the support of $\tilde{G}$. Thus the Lemma follows.
\end{proof}
\begin{Lem}\normalfont
For $\phi\in \mathcal{D}$, there exists a constant $\kappa_{3}>0$ such that for any $t\in\mathbb{Z}\setminus\{0\}$,
\begin{equation}
\text{Im}\langle (U_{0}-U)U_{0}^{t-1}\phi, U_{0}^{t}\phi\rangle \ge \displaystyle\frac{1}{2}(1+|2t|)^{-\gamma}(1-|a|^{2})\|\phi\|^{2}-\kappa_{3}t^{-2}.
\end{equation}
\end{Lem}
\begin{proof}
It follows that
\begin{equation*}
\begin{aligned}
&\hspace{5mm}\text{Im}\langle (U_{0}-U)U_{0}^{t-1}\phi, U_{0}^{t}\phi\rangle
\\
&=\displaystyle\sum_{|x|<2t}\sin(1+|x|)^{-\gamma}\|(U_{0}^{t-1}\phi)(x)\|^{2}_{\mathbb{C}^{2}}+\displaystyle\sum_{|x|\ge 2t}\sin(1+|x|)^{-\gamma}\|(U_{0}^{t-1}\phi)(x)\|^{2}_{\mathbb{C}^{2}}
\\
&\ge\displaystyle\frac{1}{2}(1+|2t|)^{-\gamma}\|\phi\|^{2}+\displaystyle\sum_{|x|\ge 2t}\Big(\sin (1+|x|)^{-\gamma}-\sin (1+|2t|)^{-\gamma}\Big)\|(U_{0}^{t-1}\phi)(x)\|_{\mathbb{C}^{2}}^{2}
\\
&\ge \displaystyle\frac{1}{2}(1+|2t|)^{-\gamma}\|\phi\|^{2}-(1+|2t|)^{-\gamma}\displaystyle\sum_{|x|\ge 2t}\|(U_{0}^{t-1}\phi)(x)\|^{2}_{\mathbb{C}^{2}}
\\
&\ge \displaystyle\frac{1}{2}(1+|2t|)^{-\gamma}\|\phi\|^{2}-(1+|2t|)^{-\gamma}\displaystyle\sum_{|x|\ge 2t}\displaystyle\frac{x^{2}}{4t^{2}}\|(U_{0}^{t-1}\phi)(x)\|^{2}_{\mathbb{C}^{2}}
\\
&\ge \displaystyle\frac{1}{2}(1+|2t|)^{-\gamma}\|\phi\|^{2}-\displaystyle\frac{(1+|2t|)^{-\gamma}}{4}\Big\|\displaystyle\frac{Q(t)}{t}U_{0}^{-1}\phi\Big\|^{2}.
\end{aligned}
\end{equation*}
Since $U_{0}^{-1}\mathcal{D}\subset\mathcal{D}$, we can apply Lemma 3.1. Hence it follows that
\begin{equation*}
\begin{aligned}
&\hspace{5mm}\text{Im}\langle (U_{0}-U)U_{0}^{t-1}\phi, U_{0}^{t}\phi\rangle
\\
&\ge \displaystyle\frac{1}{2}(1+|2t|)^{-\gamma}\|\phi\|^{2}-\displaystyle\frac{(1+|2t|)^{-\gamma}}{2}\Big\|\displaystyle\frac{Q(t)}{t}U_{0}^{-1}\phi-V_{0}U^{-1}_{0}\phi\Big\|^{2}-\displaystyle\frac{(1+|2t|)^{-\gamma}}{2}\|V_{0}U_{0}^{-1}\phi\|^{2}
\\
&\ge \displaystyle\frac{1}{2}(1+|2t|)^{-\gamma}(1-|a|^{2})\|\phi\|^{2}-\displaystyle\frac{\kappa_{1}^{2}}{2}t^{-2},
\end{aligned}
\end{equation*}
where we used that $U_{0}^{-1}V_{0}=V_{0}U_{0}^{-1}$ and Proposition 3.1. By setting $\kappa_{3}:=\kappa_{1}^{2}/2$, we have a desired inequality.
\end{proof}
In what follows, we set $\epsilon$ as $0<\epsilon<|a|/6$. We choose $G_{\epsilon}\in C_{0}^{\infty}(\mathbb{R})$ such that $0\le G_{\epsilon}\le 1$, $G_{\epsilon}(s)=1$ if $|s|\le 2\epsilon$ and $G_{\epsilon}(s)=0$ if $|s|\ge 3\epsilon$.
\begin{Lem}\normalfont
For any $\phi\in\mathcal{D}$, there exists a constant $\kappa_{4}>0$ such that for any $t\in\mathbb{Z}\setminus\{0\}$, 
\begin{equation}
\|(U_{0}-U)U^{t-1}_{0}\phi\|\le \kappa_{4}(1+|2t|)^{-\gamma} +2\|G_{\epsilon}(V_{0})\phi\|.
\end{equation}
\end{Lem}
\begin{proof} It follows that
\begin{equation*}
\begin{aligned}
&\hspace{5mm}\|(U_{0}-U)U_{0}^{t-1}\phi\|^{2}
\\
&=\displaystyle\sum_{|x|<2t\epsilon}|1-e^{i(1+|x|)^{-\gamma}}|^{2}\|(U_{0}^{t-1}\phi)(x)\|^{2}_{\mathbb{C}^{2}}+\displaystyle\sum_{|x|\ge 2t\epsilon }|1-e^{i(1+|x|)^{-\gamma}}|^{2}\|(U_{0}^{t-1}\phi)(x)\|^{2}_{\mathbb{C}^{2}}
\\
&\le 2\displaystyle\sum_{|x|<2t\epsilon}\|(U_{0}^{t-1}\phi)(x)\|^{2}_{\mathbb{C}^{2}}+(1+|2t\epsilon|)^{-2\gamma}\displaystyle\sum_{|x|\ge 2t\epsilon }\|(U_{0}^{t-1}\phi)(x)\|^{2}
\\
&\le 2\|E_{Q/t}((-2\epsilon, 2\epsilon))U_{0}^{t-1}\phi\|^{2}+\epsilon^{-2\gamma}(1+|2t|)^{-2\gamma}\|\phi\|^{2},
\end{aligned}
\end{equation*}
where for a self-adjoint operator $A$, $E_{A}(\cdot)$ is the spectral measure of $A$. Since $\|E_{Q/t}((-2\epsilon, 2\epsilon))\phi\|\le \|G_{\epsilon}(Q/t)\phi\|$, it follows that 
\begin{equation*}
\begin{aligned}
&\hspace{5mm}\|(U_{0}-U)U_{0}^{t-1}\phi\|^{2}
\\
&\le 2\|G_{\epsilon}(Q(t)/t)U_{0}^{-1}\phi\|^{2}+\epsilon^{-2\gamma}(1+|2t|)^{-2\gamma}\|\phi\|^{2}
\\
&\le 4\|G_{\epsilon}(Q(t)/t)U_{0}^{-1}\phi-G_{\epsilon}(V_{0})U_{0}^{-1}\phi\|^{2}+4\|G_{\epsilon}(V_{0})\phi\|^{2}+\epsilon^{-2\gamma}(1+|2t|)^{-2\gamma}\|\phi\|^{2}
\\
&\le 4\kappa_{2}^{2}t^{-2}+\epsilon^{-2\gamma}(1+|2t|)^{-2\gamma}\|\phi\|^{2}+4\|G_{\epsilon}(V_{0})\phi\|^{2},
\end{aligned}
\end{equation*}
where we used Lemma 3.3 in the last inequality. We note that for any $t\in\mathbb{Z}\setminus\{0\}$, $t^{-2}\le 9(1+|2t|)^{-2\gamma}$ follows. Hence it is seen that
\begin{equation*}
\|(U_{0}-U)U_{0}^{t-1}\phi\|^{2}\le\Big(36\kappa_{2}^{2}+\epsilon^{-2\gamma}\|\phi\|^{2}\Big)(1+|2t|)^{-2\gamma}+4\|G_{\epsilon}(V_{0})\phi\|^{2}.
\end{equation*}
We choose $\kappa_{4}$ as $\kappa_{4}:=(36\kappa_{2}^{2}+\epsilon^{-2\gamma}\|\phi\|^{2})^{1/2}$. Then the lemma follows.
\end{proof}
\begin{proof}[Proof of Theorem 2.1 ($0<|a|<1$)]
We take $0\neq \phi\in \mathcal{D}$ such that $E_{V_{0}}((-3\epsilon, 3\epsilon))\phi=0$. Then  $G_{\epsilon}(V_{0})\phi=0$. We only consider the case $t\rightarrow\infty$. We suppose that the limit $\phi_{+}=\lim_{t\rightarrow\infty}U^{-t}U_{0}^{t}\phi$ exists. Since $\|U^{t}\phi_{+}-U_{0}^{t}\phi\|=\|\phi_{+}-U^{-t}U_{0}^{t}\phi\|\rightarrow0$ (as $t\rightarrow\infty$), we can take $N\in\mathbb{N}$ so that $\|U^{t}\phi_{+}-U_{0}^{t}\phi\|\le (1-|a|^{2})(4\kappa_{4})^{-1}\|\phi\|^{2}$ if $t\ge  N$. 
For $t_{2}>t_{1}>N$, an application of Lemma 3.4 and Lemma 3.5 yields that
\begin{equation*}
\begin{aligned}
&\hspace{3mm}\text{Im}\langle \{W(t_{2})-W(t_{1})\}\phi, \phi_{+}\rangle
\\
&=\displaystyle\sum_{t=t_{1}+1}^{t_{2}}\text{Im}\langle (U_{0}-U)U_{0}^{t-1}\phi, U_{0}^{t}\phi_{+}\rangle+\displaystyle\sum_{t=t_{1}+1}^{t_{2}}\text{Im}\langle (U_{0}-U)U_{0}^{t-1}\phi, U^{t}\phi_{+}-U_{0}^{t}\phi\rangle
\\
&\ge\displaystyle\sum_{t=t_{1}+1}^{t_{2}}\Big\{\displaystyle\frac{1}{2}(1+|2t|)^{-\gamma}(1-|a|^{2})\|\phi\|^{2}-\kappa_{3}t^{-2}\Big\}
\\
&\hspace{20mm}-\|U^{t}\phi_{+}-U_{0}^{t}\phi\|\displaystyle\sum_{t=t_{1}+1}^{t_{2}}\big\{\kappa_{4}(1+|2t|)^{-\gamma} +2\|G_{\epsilon}(V_{0})\phi\|\big\}
\\
&\ge \displaystyle\frac{1}{4}(1-|a|^{2})\|\phi\|^{2}\displaystyle\sum_{t=t_{1}+1}^{t_{2}}(1+|2t|)^{-\gamma}-\kappa_{3}\displaystyle\sum_{t=t_{1}+1}^{t_{2}}t^{-2}
\\
&\rightarrow\infty\hspace{5mm}(\text{as }t_{2}\rightarrow\infty).
\end{aligned}
\end{equation*}
On the other hand, $\text{Im}\langle\{W(t_{2})-W(t_{1})\}\phi, \phi_{+}\rangle$ is bounded by $2\|\phi\|^{2}$. This is a contradiction.
\end{proof}

\end{document}